\documentclass[twocolumn,showpacs,preprintnumbers,amsmath,amssymb]{revtex4-1}
\usepackage{amsmath}
\usepackage{graphicx}
\usepackage{dcolumn}
\usepackage{bm}
\usepackage{natbib}
\makeatother

\begin{document}

\title{Stimulated emission of fast Alfv{\'e}n waves within magnetically confined
fusion plasmas}

\author{J W S Cook$^{1,2}$, R O Dendy$^{3,2}$, and S C Chapman$^{2}$}

\affiliation{$^{1}$First Light Fusion Ltd., Unit 10, Oxford Industrial Park,
Yarnton, OX5 1QU, U.K.}

\affiliation{$^{2}$Centre for Fusion, Space and Astrophysics, Department of Physics, Warwick University, Coventry CV4 7AL, U.K.}

\affiliation{$^{3}$CCFE, Culham Science Centre, Abingdon, Oxfordshire OX14 3DB, U.K.}

\begin{abstract}
A fast Alfv{\'e}n wave with finite amplitude is shown to grow by a stimulated 
emission process that we propose for exploitation in toroidal magnetically 
confined fusion plasmas. Stimulated emission occurs while the wave propagates 
inward through the outer mid-plane plasma, where a population inversion 
of the energy distribution of fusion-born ions is observed to arise naturally.
Fully nonlinear first principles simulations, which self-consistently evolve
particles and fields under the Maxwell-Lorentz system, demonstrate this novel 
``alpha-particle channelling'' scenario for the first time.
\end{abstract}

\maketitle
There is overwhelming evidence for the natural occurrence of a substantial
inversion in the energy distribution of fusion-born ions, and of other
energetic minority ions, at the outer mid-plane edge of large tokamak
and stellarator plasmas. Spontaneous relaxation of this population
is observed in the form of intense ($\sim\negmedspace10^{4}$ times
black-body) suprathermal ion cyclotron emission (ICE), resulting from
excitation of waves on the fast Alfv{\'e}n branch by means of the collective
magnetoacoustic cyclotron instability (MCI) \cite{ref:belikov1976sovphystechphys,ref:dendy1992physfluidsb,ref:dendy1993physfluidsb,ref:dendy1994physplas,ref:dendy1994physplas2,ref:mcclements1996physplas,ref:dendy1995nuclfus,ref:gorelenkov1995nuclfus,ref:fulop1998nuclfus}.
These waves are primarily electromagnetic, with an electrostatic component,
and are observed at narrow spectral peaks at frequencies of tens or
hundreds of MHz, corresponding to local sequential ion cyclotron harmonics.
For the unique deuterium-tritium plasmas in JET and TFTR \cite{ref:cottrell1988prl,ref:schild1989nuclfus,ref:cottrell1993nuclfus,ref:cauffman1995nuclfus,ref:mcclements1999prl},
the energy-inverted ion population driving the MCI comprises a subset
of the centrally born trapped fusion products, lying just inside the
trapped-passing boundary in velocity space, whose drift orbits make
large radial excursions to the outer mid-plane edge. In other large
magnetically confined fusion (MCF) plasmas, the corresponding energetic
ion population arises either from neutral beam injection (NBI) or
from trace fusion reactions, or both. For example, in the edge plasma
of the ASDEX-Upgrade \cite{ref:dinca2011epsconfprof,ref:dinca2013thesis}
and JT-60U \cite{ref:ichimura2008nuclfus,ref:sato2010plasfusres} tokamaks,
ICE is detected at the cyclotron harmonics of the proton, triton and
$^3$He products of fusion reactions in pure deuterium plasmas. ICE is
also used as a diagnostic of lost fast ions in the DIII-D tokamak \cite{ref:watson2003revsciinstr,ref:heidbrink2011ppcf}
and the large stellarator-heliotron LHD \cite{ref:saito2013}. Recent
first-principles simulations \cite{ref:cook2013ppcf,ref:carbajal2014physplas}
exploiting particle-in-cell and hybrid kinetic-fluid codes indicate
that observed ICE can grow from noise into the nonlinear saturated
regime on the fast timescales of relevance. These simulations further
confirm the physics assumptions made in analytical studies \cite{ref:dendy1992physfluidsb,
  ref:dendy1993physfluidsb,ref:dendy1994physplas,ref:dendy1994physplas2,ref:mcclements1996physplas,
  ref:dendy1995nuclfus,ref:gorelenkov1995nuclfus,ref:fulop1998nuclfus}
of the MCI for ICE interpretation. Thus the combination of high temperature
plasma conditions with toroidal magnetic confinement geometry leads
naturally to a spatially localised source of spontaneous collective
emission of spectrally structured electromagnetic waves with strongly
suprathermal intensity.
\par In this paper we propose and investigate, for the first time, 
a stimulated emission counterpart to the observed spontaneous emission
process. We explore the behaviour of finite-amplitude fast Alfv{\'e}n
waves that are incident on plasma which contains, in addition to majority
thermal ions and electrons, a small minority energetic
particle population that has an inversion in its energy distribution.
This inversion corresponds to that in the ICE-generating population
in tokamaks and stellarators. For a case where the initial amplitude
of the incident fast Alfv{\'e}n wave corresponds to a field energy density
which is only $\sim0.2$\% of the kinetic energy density of the minority
ion population, the simulations reported below show that up to $\sim15$\%
of the energy stored in the minority ion population can be transferred
to the wave, on fast timescales between two and
five ion cyclotron periods, for normalised wave energies in the range
$10^{-2}$ to $10^{-8}$. The magnitude of the eventual increase in
wave energy, normalised to the kinetic energy of the minority ions,
is found to be independent of the imposed wave energy. 

These proof-of-principle results indicate that stimulated emission
of inward propagating fast Alfv{\'e}n waves in the edge region may 
become a significant addition to the techniques for alpha-channelling in 
MCF plasmas. Alpha-channelling \cite{ref:fisch1992prl, ref:fisch1995physplas}
denotes the exploitation of the free energy in fusion-born ions,
for example to: drive internal currents \cite{ref:cook2010prl, 
ref:cook2011ppcf, ref:ochs2015physplas}; 
radially transport and cool resonant particles
 \cite{ref:fisch1992prl, ref:fetterman2008prl}; 
preferentially heat fuel ions \cite{ref:zhmoginov2011prl, ref:clark2000physplas}.
This is in contrast to conventional alpha-particle heating
of the thermal plasma by electron collisions. A variety of alpha-channelling
mechanisms have been studied, typically involving (as in the novel
case proposed here) fast collective relaxation and radiation. There
are various potential alpha-channelling applications of a fast Alfv{\'e}n
wave originating from stimulated emission by fusion-born ions at the
outer mid-plane edge of MCF plasmas. The additional deuteron kinetic energy
is initially in the form of coherent oscillition supporting the 
stimulated fast Alfv{\'e}n wave; this energy can be thermalised through
collisions. Also ion cyclotron resonant damping of the amplified wave
deeper within the plasma would by-pass the lossy electron channel. 
This process could also 
extend the use of ICE as a diagnostic for fusion-born ion populations
in future deuterium-tritium plasmas in JET and ITER, as proposed in
Refs. \cite{ref:dendy2015ppcf,ref:mcclements2015nuclfus}.
For maximum effect, it might also be desirable
to nudge fusion-born alpha-particles in the plasma core onto drift
orbits which place them in the emitting population. 
ICE occurs in solar-terrestrial and, probably, astrophysical
plasmas \cite{ref:mcclements1993jgr,ref:dendy1993jgr,ref:rekaa2014journalanme,
ref:posch2015jgeores}, suggesting that stimulated emission of fast
Alfv{\'e}n waves, by the mechanism investigated here, could also arise
in those natural plasma contexts.

We deploy a first principles fully-kinetic relativistic 1D3V Maxwell-Vlasov
particle-in-cell (PIC) code EPOCH \cite{ref:arber2015ppcf}
to self-consistently integrate the electromagnetic field simultaneously
with the full distributions of 1 keV electrons, majority thermal 1
keV deuterons and minority energetic 3.5 MeV alpha-particles. A uniform
magnetic field of 2.1 T is initialised along the $z$-axis, whilst
the spatial domain is aligned with the $x$-axis.
All species are initially spatially homogeneous, and the electron
density is $10^{19}$ $m^{-3}$. The initial alpha-particle distribution
is a ring in velocity space, $f(v_{\perp}, v_{\parallel})=\delta(v_{\perp}-u_{\perp})\delta(v_{\parallel})$,
where $v_{\parallel}\equiv v_{z}$, $v_{\perp}^{2}=v_{x}^{2}+v_{y}^{2}$,
and $u_{\perp}$ corresponds to the speed of 3.5 MeV. The ratio of
the alpha-particle population density to the deuteron density $\xi$
is $10^{-3}$. This is an order of magnitude higher
than in experiments \cite{ref:cottrell1988prl, ref:cottrell1993nuclfus}, 
and is chosen raise the
signal above the noise in these periodic simulations. Each species
is represented by 200 particles per cell, with a total of 4096 cells.
Presented below are results from a parameter scan in energy of an
imposed fast Alfv{\'e}n wave. This energy is varied from 1\% of that initially
in the alpha-particle population in steps of factors of 10 down to
$10^{-8}$. In this preliminary study, the fast Alfv{\'e}n wave has a frequency
of $18\omega_{c\alpha}$, and we set the simulation
domain length to 120 wavelengths. This frequency is one at which
the MCI grows in the absence of any imposed waves. Perturbed quantities
and the wavenumber are determined by the cold plasma dispersion relation \cite{ref:stix1992book}
at this frequency. The imposed wave perturbs the electric field components such that:

\begin{equation}
\left[\begin{array}{cc}
S-n^{2} & -iD\\
iD & S-n^{2}
\end{array}\right]\left[\begin{array}{c}
E_{x}\\
E_{y}
\end{array}\right]=\left[\begin{array}{c}
0\\
0
\end{array}\right],
\end{equation}
where $S$, $D$ and the perpendicular refractive index $n$ are defined in 
Ref. \cite{ref:stix1992book}. This linear cold plasma
wave is an approximation to the hot nonlinear plasma wave represented
in the fully kinetic PIC code that we deploy \cite{ref:arber2015ppcf}. 
The wave is initialised in the simulation with coherent spatial perturbations to: the
$B_z$, $E_x$, and $E_y$ fields; the $J_x$ and $J_y$ current components; 
the $v_x$ and $v_y$ components of the velocity of electrons and deuterons;  
and the electron and deuteron densities. The alpha-particles are initially
unperturbed. Maxwell's equations and the conservation of mass and 
momentum determine the other perturbed quantities at 
initialisation time. Thus the imposed wave is a self-consistent solution of
the system of equations solved by the PIC code and
is initially supported by the majority deuteron and electron populations
only. Thereafter the evolution of the perturbed fields and particles, including 
alpha-particles, is governed self-consistently by the PIC solver. 
These simulations present an idealised scenario where
energetic alpha-particles initially have: an infinitely narrow distribution
function; no spatial inhomogeneities; and no particle sources
or sinks. 

Electrons and deteurons are initialised with
a quiet start method employed in Refs. \cite{ref:cook2011ppcf,ref:cook2013ppcf}
The computational macro-particles are initialised uniformly
in configuration space and randomly in velocity space so that their
distributions approximate the initial conditions. 
The cell-integrated first and second moments of the
distribution functions are used to correct the particle velocities
to match the initial conditions exactly on a cell averaged basis.
The alpha-particles are initialised uniformly in configuration space,
and in each cell they are positioned in gyro-angle such that spatially
adjacent particles have opposing velocities, and are offset in gyro-angle
by the same random amount. On a cell averaged basis, charge density
is uniform and the cell summed current is zero, hence our initial
conditions are satisfied exactly. 

\begin{figure}
\includegraphics{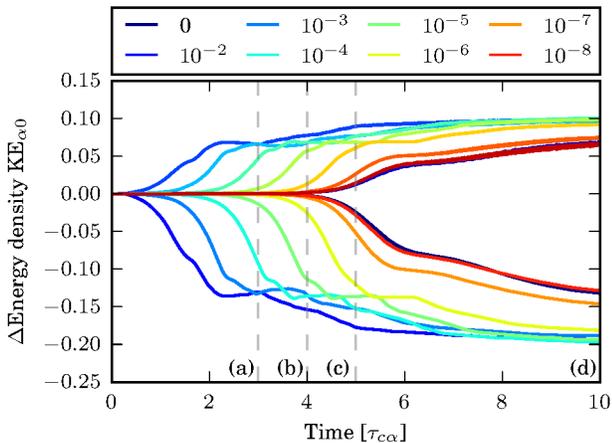}
\caption{
\label{fig:alpha_deuteron_energy}
(colour online). Stimulated emission of 5\% to 10\%
of the energy of the alpha-particle population over a few gyroperiods,
when subjected to resonant fast Alfv{\'e}n waves that have much lower
energy density. Temporal evolution of the minority alpha-particle
population energy density (traces below $\Delta\mathrm{Energy}=0$)
and deuteron energy density (traces above $\Delta\mathrm{Energy}=0$)
in response to applied waves with a range of energy densities. Pairs
of traces for alpha-particles and deuterons show energy
change in the presence of waves with energy densities ranging between
$10^{-2}$ and $10^{-8}$ that of the energy initially in the alpha-particles,
and the null case without an imposed wave. Energetic minority alpha-particles
transfer energy to thermal majority deuterons on shorter timescales
when subjected to higher amplitude fast Alfv{\'e}n waves. 
The eventual level of energy transfer is broadly the same. Vertical
dashed lines annotated (a)-(d) indicate snapshots in time displayed
in Fig. \ref{fig:alpha_vx_vy}. Change in energy density (ordinate)
is plotted in units of the initial alpha-particle energy density and
time (abscissa) is plotted in units of the alpha-particle cyclotron
time period $\tau_{c\alpha}=2\pi/\omega_{c\alpha}$.
}
\end{figure}

Figure \ref{fig:alpha_deuteron_energy} shows the temporal evolution
of the normalised change in the energy density of the
deuteron and alpha-particle populations, averaged 
across the spatial simulation domain. Increasing
the amount of energy in the applied wave causes energy to be extracted
faster from the alpha-particle population and deposited in the deuterons
whose coherent oscillations support the linearly \cite{ref:mcclements1996physplas}
and non-linearly \cite{ref:carbajal2014physplas} excited waves.
Importantly much of this energy is passed to the background deuterons
(see Fig. \ref{fig:alpha_deuteron_energy}), which represent the
fuel ions. Hence we have identified a new method for collisionless energy
transfer from fusion product alpha-particle to fusion fuel ion. Notably
the deuterons are energised but not heated; their extra kinetic energy
sustains the fast Alfv{\'e}n wave through collective oscillations
but does not simply broaden their distribution function. However, 
the collective motion would heat the fuel ions through collisions 
on a longer timescale than is considered here.

\begin{figure}
\includegraphics{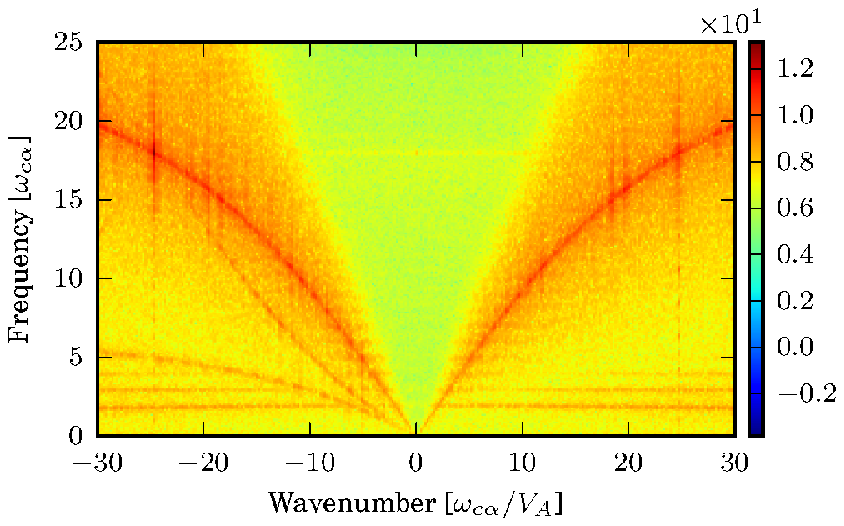}

\vspace{-1.05cm}

\includegraphics{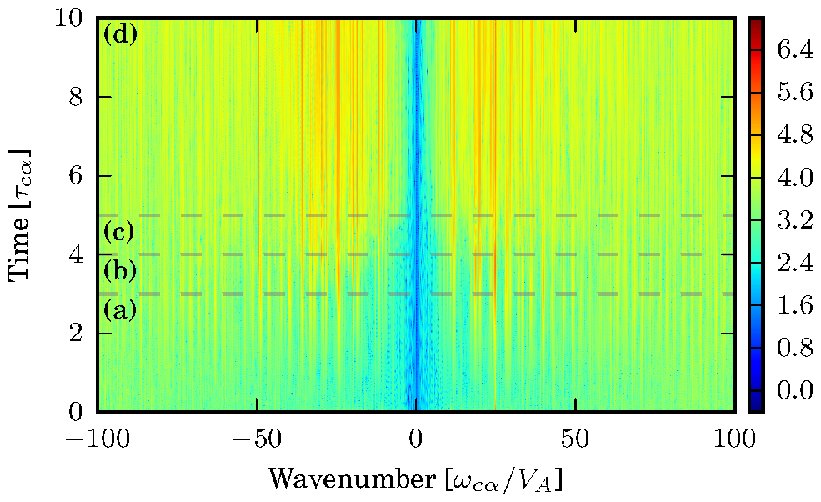}
\caption{\label{fig:wk}
(colour online). Dominant waves, and their linear and nonlinear interactions,
identified from Fourier transforms of the computed $E_{x}$ field
in the simulations. These are obtained by computing the fast Fourier
transform of the whole spatial and temporal domain from a simulation
with an imposed Alfv{\'e}n wave with an initial energy $10^{-5}$ that
of the initial alpha-particle population. Upper panel plots $(\omega,k)$
amplitudes; lower panel, $(t,k)$ amplitudes.
}
\end{figure}

Figure \ref{fig:wk}(a) plots the power in the spatio-temporal fast
Fourier transform of the electric field component aligned with the
grid, $E_x$, integrated over the whole spatial and temporal domain. Notable
peaks in the spectrum are located at: the frequency and wavenumber
of the imposed wave $(\omega=18\;\omega_{c\alpha},\:k\simeq-25\;\omega_{c\alpha}/V_{a})$;
frequencies and wavenumbers of oppositely traveling waves to which
the imposed wave couple $(14\lesssim\omega/\omega_{c\alpha}\lesssim20,18\lesssim k\omega_{c\alpha}/V_{A}\lesssim30)$;
and the first harmonic of the imposed wave $(\omega=36\:\omega_{c\alpha},\:k\simeq-50\:\tau_{c\alpha}/V_{a})$. Figure \ref{fig:wk}(b) 
plots the power in $E_x$ Fourier transformed in space but not in time,
and integrated over the whole spatial domain. We see the onset of
nonlinearity in the forward propagating waves
$(3\lesssim t/\tau_{c\alpha}\lesssim4,18\lesssim kV_{A}/\omega_{c\alpha}\lesssim30)$,
 which couple to the imposed wave. 
Referring to Fig. \ref{fig:alpha_deuteron_energy}, this matches 
the end of the linear phase of the instability.

\begin{figure}

\includegraphics{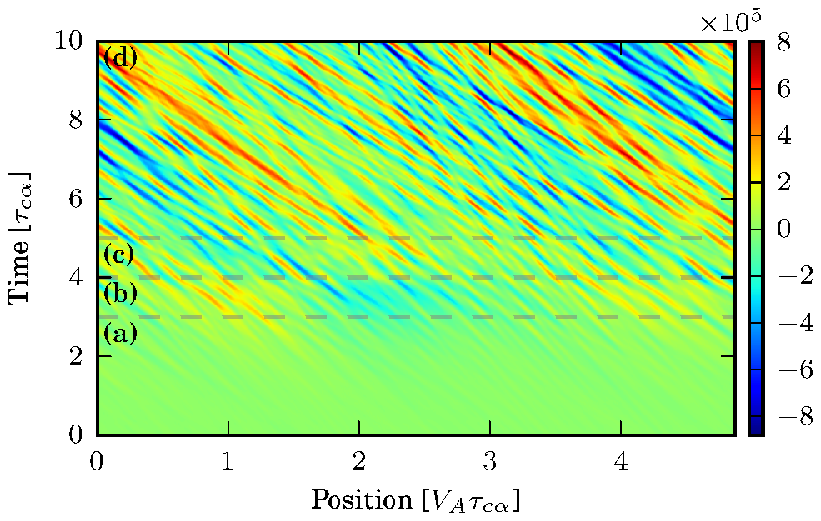}

\vspace{-1.05cm}

\includegraphics{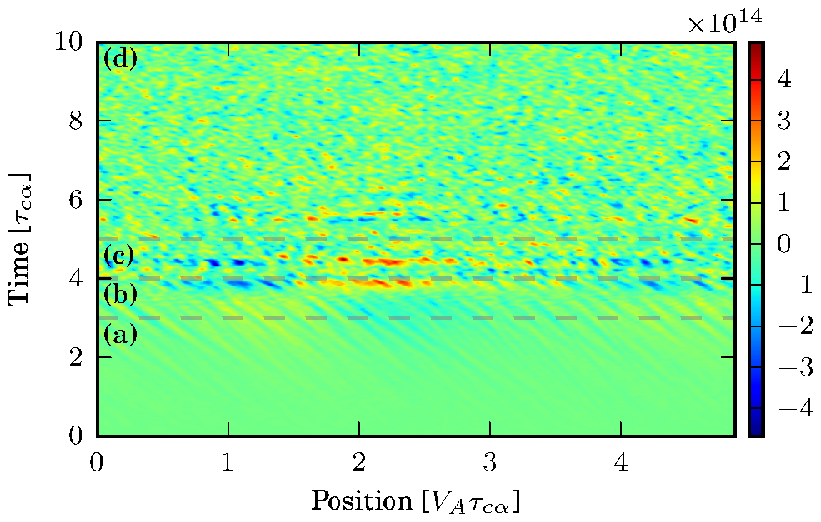}

\caption{\label{fig:tx_dft}
(colour online). Coherent field oscillations persist through, and beyond, 
the sudden transition to strong nonlinearity in alpha-particle dynamics.
Linear (early time) and nonlinear (late time) propagation
and evolution of energy wavepackets at the imposed frequency and wavenumber
$(\omega_{i},k_{i})$.
Shading indicates the spatio-temporal amplitude of
the moving windowed 2D DFT at $(\omega_{i},k_{i})$ of 
the $E_{x}$ field (upper panel) and
alpha-particle number density (lower panel),
for a window of size $2\pi(\omega_{i}^{-1},k_{i}^{-1})$.
Data are from the simulation shown in Fig. \ref{fig:wk}. 
}
\end{figure}

\begin{figure}
\includegraphics{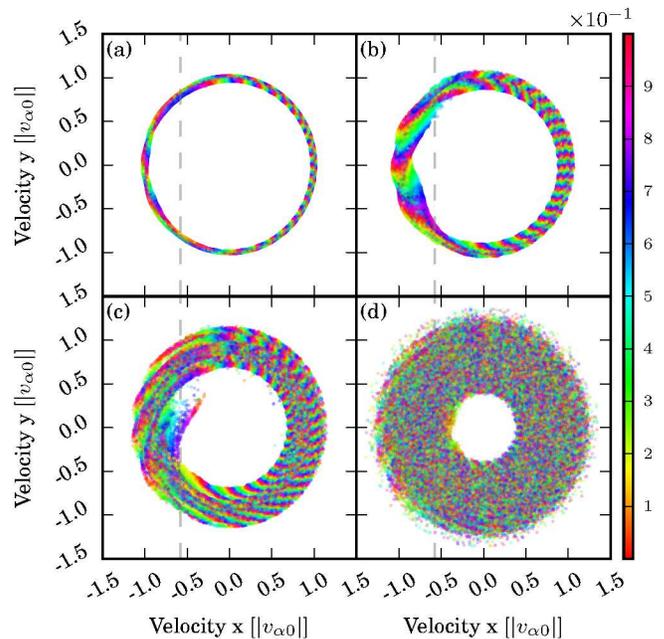}

\caption{\label{fig:alpha_vx_vy}
(colour online). Sudden transition to nonlinear alpha-particle dynamics at 
$t\simeq 4\tau_{c\alpha}$.
Linear and nonlinear wave-particle interactions
shown in the $(v_{x},v_{y})$ space of a randomly selected sample
of 10\% of alpha-particles at four times (a) to (d) which are marked
in Figs. \ref{fig:wk} and \ref{fig:tx_dft}. Shading represents
the value of $x$ mod $\lambda$ at $t\simeq3,4,5,10$ $\tau_{c\alpha}$.
The dashed trace shows the phase speed
of the imposed fast Alfv{\'e}n wave. Velocity is in units of the initial
alpha-particle speed. Data are from the same simulation shown in
Fig. \ref{fig:wk}.
}
\end{figure}

The extracted energy travels through the plasma at the wave-packet's
group speed, which in a hot plasma
is analytically inaccessible, see Eq. 69 in Ref. \cite{ref:dendy1992physfluidsb}.
We measure it in Fig. \ref{fig:tx_dft}, which shows the spatio-temporal
evolution of wave-packets of the imposed wave with frequency and wavenumber
$(\omega_{i},k_{i})$ in the $E_{x}$ field (upper panel) and alpha-particle
number density (lower panel). Amplitude at each $(t,x)$ location is
the sum of the absolute values of a discrete Fourier transform (DFT)
at $(\omega_{i},k_{i})$, with a spatio-temporal extent 
$2\pi(k_{i}^{-1},\omega_{i}^{-1})$ 
that overlaps this point. This moving DFT window is applied to the whole
$(t,x)$ domain: periodically in the spatial direction, and aperiodically
in the temporal. The upper panel indicates that the energy
in the imposed wave travels from right to left at a velocity of $-0.4\pm0.06$
$\mathrm{V_{A}}$ $\simeq0.55\pm0.08$ $\omega_{i}/k_{i}$, where
$k_{i}<0$. We infer from this the group velocity of the wave packets
of transferred energy. In a tokamak, a wave launched from
the outboard edge would be amplified by alpha-particle energy and
travel further towards the core at approximately half its phase speed.
The lower panel indicates the abrupt start, at 
$t \simeq 4\tau_{c\alpha}$, of the nonlinear phase
in the alpha-particle spatial distribution at the imposed wave's wavenumber
and frequency (c.f. the trace labelled $10^{-5}$ in Fig. \ref{fig:alpha_deuteron_energy}).

Our fully kinetic 1D3V simulations give a detailed view of the temporal
evolution of perturbations to the alpha-particle distribution function.
Figure \ref{fig:alpha_vx_vy} plots the $x$ and $y$ components
of velocity of the alpha-particles, where shading shows
the value of $x$ mod $\lambda$ where $\lambda$ is the wavelength
of the imposed wave, at four snapshots in time: panel (a) shows the
imprint of the MCI during the linear stage of the process, at 3 $\tau_{c\alpha}$;
panels (b) and (c) show the nonlinear development at 4 and 5 $\tau_{c\alpha}$
respectively; and panel (d) shows the final state at 10 $\tau_{c\alpha}$.
The imposed wave in this simulation has a negative velocity as shown
by the vertical dashed traces, and is in phase-space resonance with
alpha-particles; these wave-particle interactions are visible in the
left-right asymmetry of the distribution functions. The complex
structure in panels (b) and (c) shows progression
to the nonlinear phase.

In this paper we have identified a process whereby the effective energy
confinement of the fusion-born alpha-particle population is significantly
enhanced. As an externally applied fast Alfv{\'e}n wave of initially low
amplitude propagates inward, it is amplified by stimulated emission
of energy from the alpha-particles at the outer midplane, whose velocity
distribution is inverted. This stimulated emission arises because
this alpha-particle population is naturally linearly unstable to the
MCI at the selected $(\omega,k)$ of the applied wave. We have measured
the group velocity of the inward propagating wave to be $0.55\pm0.08$
$V_{A}$ and show that amplification occurs on a timescale $\sim2$-$5$
$\tau_{c\alpha}$. Approximately half the 5\%-10\% alpha-particle
energy released is transferred inward to the fuel ions that support
this wave, and the waves to which it couples, thereby leaving 
less free energy available for spontaneously excited waves that leave 
the plasma. The amplified wave
passes into a nonlinear regime in which the background real space-configuration
- initially uniform - becomes modulated, resulting in wavepackets
travelling at non-identical speeds. These are seen to collide, giving
rise to increased phase space complexity; this study does not 
investigate the spatial transport of alpha particles that accompanies
energy exchange. This newly identified stimulated emission process 
is an instance of alpha-particle power channelling,
which in general rests on velocity space resonance giving rise to
beneficial real space energy transport.

It is a pleasure to thank N. J. Fisch for stimulating discussions. 
This work was part-funded by the RCUK Energy Programme {[}under grant
EP/I501045{]} and the European Communities. The views and opinions
expressed herein do not necessarily reflect those of the European
Commission. The EPOCH code used in this research was developed
under UK Engineering and Physical Sciences Research Council grants
EP/G054940/1, EP/G055165/1 and EP/G056803/1.

\end{document}